\documentclass[12pt]{article}
\usepackage{graphicx}
\newcommand {\be}{\begin{equation}}
\newcommand {\ee}{\end{equation}}

\begin{document}

 \def\Colore#1#2#3{
}
 \def\Rosso{\Colore{1}{0}{0} }
 \def\Verde{\Colore{0}{1}{0} }
 \def\Blu{\Colore{0}{0.}{1} }
 \def\Bluviola{\Colore{0}{0.65}{1} }
 \def\Celeste{\Colore{0}{1}{1} }
 \def\Nero{\Colore{0}{0}{0} }
 \def\Bianco{\Colore{1}{1}{1} }
 \def\Rossoscuro{\Colore{0.80}{0.00}{0.20} }
 \def\Verdescuro{\Colore{0.00}{0.80}{0.20} }
 \def\Bluscuro{\Colore{0.20}{0.00}{0.80} }
 \def\Viola{\Colore{0.65}{0.0}{0.65} }
 \def\Lilla{\Colore{1}{0}{1} }
 \def\Bluchiaro{\Colore{0}{0.75}{1} }
 \def\Blumedio{\Colore{0}{0.5}{1} }
 \def\Bluforte{\Colore{0}{0.25}{1} }
 \def\Arancione{\Colore{1}{0.5}{0} }
 \def\Verdechiaro{\Colore{.5}{1}{0} }
 \def\Marrone{\Colore{.6}{.4}{0.2} }
 \def\Giallo{\Colore{1}{1}{0} }
 \def\Gialloscuro{\Colore{.8}{.7}{0} }
 \def\Rossochiaro{\Colore{1}{0}{0.5} }
 \def\Rosa{\Colore{1}{0.5}{.5} }
 \def\Grigio{\Colore{0.5}{0.5}{0.5} }
  \def\Blut{\Bluforte}
\baselineskip 18pt

\title {Statistics of Poincar\' e recurrences for a class of smooth circle maps}

\author { Nikola Buric $^1$, Aldo Rampioni $^2$ and  Giorgio Turchetti $^2$
\\ (1) Department of Physics and Mathematics,
\\ Faculty of Pharmacy, University of Beograd,
 \\ Vojvode Stepe 450, \\11000 Beograd, Yugoslavia.
\\ (2) Department of Physics, University of Bologna,
\\ INFN Sezione di Bologna, Italy.
 }

\date{\today}

\maketitle {}

\begin{abstract}

Statistics of Poincar\' e recurrence for a class of circle maps,
 including sub-critical, critical, and super-critical
  cases, are studied. It is shown how the topological
  differences in the
 various types of  the dynamics are  manifested in the
 statistics of the
 return times.

\end{abstract}

PACS:

\newpage
\noindent
\section{  Introduction}

Statistics of the Poincar\' e recurrences, i.e. return times statistics,
 has recently gained renewed importance in the theory of dynamical
 systems, primarily due to the fact that it could be used as an
 indicator of the statistical properties of the system's global
 dynamics on large parts of its phase space (see for example
  \cite{Chir99},\cite{Vaienti1}).
 For instance, the first return time can be used to calculate
 the metric entropy of a system with an ergodic invariant measure
 \cite{Ors}, and seems to be related to other generalized dimensions
  used to describe the fractal properties of the dynamics, at least
  for certain types of dynamical systems \cite{Vaienti2}.
In this paper we study the Poincar\' e recurrences for maps of the
circle,  that can display quasi-periodic, bi-stable or chaotic
dynamics, depending on the
 values of  their two parameters.


Given  a discrete dynamical system  on a phase space
 $M$  with  a transformation $T: M \to M$  and a reference measure
 $\mu$ on $M$, the   first return time,    in a measurable set
 $A\subset M$,   of a point $x\in A$,  is defined by
\be
\tau_A(x)=\inf_k \{ T^k(x)\in A\}.
 \ee

The first return time  $\tau_A$ for the set $A$, and the average
 return time $<\tau_A>$ for the set $A$,  are the following
\be
 \tau_A=\inf_{x\in A} \tau_A(x),\qquad  <\tau_A>=\int_A
\tau_A(x) d\mu_A(x),
 \ee where $\mu_A(x)$ is the
conditional   measure $\mu_A(B)=\mu(B)/\mu(A)$ for any $B\subseteq
A$.

  With $<\tau_A>$ and $\tau_A(x)$ we define the probability  $F_A(t)$  that the
normalized  first return time  in the set $A$ is larger than $t$
for the points in the set $A$
 \be
 F_A(t)=\mu_A(A_{>t})
\qquad\qquad A_{>t}\equiv\left \{x \in A_t:\
 {\tau_A(x)/ \langle \tau_A\rangle } > t \right\}.
\ee
 A limit probability measure $ F_x(t)$ may be associated
to any point $x$ by considering a sequence of
 neighborhoods $A_k$ of $x$ with $\mu (A_k) \to 0$ as
 $k \to \infty$. For ergodic systems the limit is the
 the same for almost every point and in
  this case the average return
 time in a domain $A$ is given by
\be
 \langle \tau_A \rangle ={1\over \mu(A)}
\ee
 according to the well known Kac's lemma. For some classes of
hyperbolic dynamical systems \cite{Hir95} it has been proved that
the return times spectrum follows the exponential-one decay law
$F_x(t)=e^{-t}$ at almost every $x$. If $x$ belongs to the  dense
set of the unstable periodic points  then $ F_x(t)=a\exp(-\alpha
t)$ where $a$ and $\alpha$ depend on the period.
     It is known that the properties of the distribution of the return times
can give criteria for the existence of an equilibrium and the rates of mixing \cite{Young}.
 A  polynomial  decay law   for  $F_x(t)$  was found for
 integrable area-preserving  maps \cite{Aldo}\cite{Hu}.
On the basis of numerical  computations, the polynomial decay of
return times spectra  $F_x(t)$ was  also claimed   to be  a
generic property  of Hamiltonian systems with mixed phase space,
where complicated self similar fractal structures   are present
\cite{Zavlavsky},\cite{Chir99}. In fact, for such systems, a
convex combination of the
 power law and the exponential law decays seems to provide a very good fit of
 $F_x(t)$ and is theoretically justified \cite{us2002}.

Our aim is to study the statistics of the first return times for
smooth perturbations (non necessarily invertible) of the uniform
 rotations $T_{k,\Omega}$ on the circle $S^1$, \be
 \theta\rightarrow
 T_{k,\Omega}(\theta)=\theta+\Omega+ {k\over
2\pi}f(\theta),
 \ee
 where $\theta\in S^1$  and $k\in R^+$ and $\Omega\in R^+$
are  the   parameters of the map. The function $f(\theta)$ is  a
trigonometric polynomial such that the maps are monotonic for
$k\in [0,1)$ and  non-invertible for $k>1$.
 The details of the dynamics and the structure of the
 bifurcation diagram have been thoroughly studied for the sine-circle map
 given by $ T_{k,\Omega}(\theta)=\theta+\Omega+{k\over 2\pi}\sin{2\pi \theta}$
 (see for example \cite{Denj}- \cite{Khanin3}). Other families
 of the form (4) are used to study the universality of the properties
  found for  the sine-circle map.

 The main results of our analysis of  the recurrence times for
the circle maps of the form (5) can be summarized as follows.
 For   $k<1$, where  the map is invertible  and
diffeomorphic to a rotation,   if the rotation number is
diophantine, three return times are  observed. This is in
agreement with Slater's theorem \cite{Slater}, which we extend
from linear rotations to the diffeomorphisms of the circle, see
section 2. At the critical value $k=1$ three return times are
still observed and  the average return times  allow an effective
reconstruction of the invariant measure, see section 3. In the
super-critical case (section 4), we show the appearance of a
continuous spectrum  $F_{\theta}(t)$ for maps that have chaotic
orbits at $k=1+\epsilon$ (section 4.1), and for maps
$T_{k,\Omega}$ for $k$ beyond the
 corresponding accumulation point of period-doubling bifurcations (section 4.3).
Also, the properties of the return times could be used to detect
the existence of attracting periodic orbits (section 4.2).

 Our conclusion is that spectrum reflects, the
bifurcations in the topological properties of the dynamics. and
is a useful tool to investigate them.


\section{Dynamical properties of the circle map}

Let us briefly recapitulate some of the properties of the
 circle maps (5). For our purposes, we
distinguish three regions depending on the parameter $k$: the
sub-critical region $k<1$, the weakly super-critical region
$k=1+\epsilon$  and the strongly super-critical region $k>>1$ .

\subsection{Sub-critical and  critical region}

For  $k\leq 1$ the circle  map $T_{k,\Omega}$  is an orientation
preserving homeomorphism  of the circle, and for $k<1$ the map is
a diffeomorphism. In any case, its topological properties are
fixed by the rotation number $\omega$ defined by \be
\omega=\lim_{n\rightarrow\infty} {\bar T^n(\theta)\over n}, \ee
where $\bar T$ is the lift of $T$ on the real line, and
$0\leq\omega <1$  for the definition to be unique.

 The map $T_{k,\Omega}$, for $k<1$ is conjugate to the linear
 rotation by the  angle $\omega(k,\Omega)$.
  The
properties of the conjugation $\Theta=\Phi_{k,\Omega}(\theta)$
with the linear  rotation  $R_\omega: \Theta\to \Theta+\omega$
depend on the arithmetic properties of $\omega$. For  a generic
rotation number $\Phi_{k,\Omega}$ is a homeomorphism according to
Denjoy's theorem  \cite{Denj}.

Furthermore if $\omega$ is a Diophantine irrational the
conjugation $\Phi_{k,\Omega}$ is a diffeomorphism \cite{Herman}.
In such a case   the topological and metric properties  of the
linear rotation  $R_\omega$  extend to the map $T_{k,\Omega}$
by using \be
 T^n_{k,\Omega}= \Phi^{-1}_{k,\Omega}\circ  R^n_\omega \circ
\Phi_{k,\Omega} \ee Slater's theorem \cite{Slater}, stating
that for any irrational linear rotation, and any connected
interval  there are at most three different return times, one of
them being the sum of the others, extends  from linear rotations
to the map $T_{k,\Omega}$. Two of the three return times  are
always the consecutive denominators in the continued fraction
expansion of the irrational rotation number $\omega$, and one of the
 return times is a sum of the other two. Two points
$\theta, \;T^n_{k,\Omega}(\theta)$  in a connected interval $A$
are mapped into $\Theta=\Phi_{k,\Omega}(\theta)$ and $R_\omega^n
(\Theta)$ of the connected interval $\Phi_{k,\Omega}(A)$

A critical map $T_{k=1,\Omega}$ is only a homemorphism of the
circle but is still characterized by a unique rotation number.
 Our calculations indicate that, at least, for sufficiently irrational
  rotation numbers
  there are still only three return times, like in the
  sub-critical case.

\subsection{Weakly super-critical region}

When $k>1$ the map ceases to be invertible and  it does not have a
unique rotation number.
 For a given  $(k>1,\Omega)$
 quasi-periodic, chaotic, at most two stable periodic
 orbits, and orbits asymptotic to the latter can coexist.
  The points in the parameters plane $(k,\Omega)$ that
  correspond to sub-critical maps with
   rational $P/Q$ rotation numbers form domains,
   the $P/Q$ tongues, that can be extended
   above the critical line $k=1$ where they  start to overlap.
   The boundaries of the tongues, correspond to tangent bifurcation, and
    can be found by solving for $\theta$ and $\Omega$ the equations
   \be
   T^{(Q)}_{k,\Omega}(\theta)=\theta+P,\qquad
   {T^{(Q)}_{k,\Omega}(\theta)}^{'}(\theta)=1,
   \ee
 where $T^{(Q)}_{k,\Omega}(\theta)$ is the $Q$-th application of the map,
 ${T^{(Q)}_{k,\Omega}(\theta)}^{'}$ its derivative (see for example \cite{Bohr2}).

   The union of all $P/Q$ tongues at $k=1$ has full measure, and
   for $k=1+\epsilon$ any map $(k,\Omega)$ belongs to intersection
   of two tongues $P_1/Q_1$ and $P_2/Q_2$ that correspond to Farey neighbours
    with sufficiently high
   $Q_1$ and $Q_2$. Actually, $(k,\Omega)$ is in the intersection of
  the tongues that correspond to all rationals deeper in the Farey
  tree and between $P_1/Q_1$ and $P_2/Q_2$. To predict the
  dynamics of the map one has to  know a very complicated fine
  structure of bifurcations inside the overlapping tongues.
    The lines in the $(k<1,\Omega)$ plane that correspond to maps with
irrational rotation numbers become, for $k>1$, domains such that
a map in such a domain have at least one orbit with the
corresponding irrational rotation number, but also has orbits
with other rotation numbers consistent with the $P/Q$ tongues that
overlap, and chaotic orbits \cite{Boyland}. Weakly super-critical
maps of the form (5) with chaotic orbits have been used to study
the quasi-periodic rout to chaotic dynamics
\cite{Bohr1}, \cite{MacKay},\cite{Rand}.

\subsection{ Strongly  super-critical region}

Besides the maps $(k,\Omega)$ with
the small $k>1$ that have chaotic orbits, there are such $\Omega$
that maps $(k,\Omega)$ inside a $P/Q$ tongue will have strongly
chaotic behaviour only if  $k>K_c(P/Q)$, where $K_c(P/Q)$ is the
value of $k$ at which the period doubling bifurcations inside
$P/Q$ tongue accumulate, which could be quite
 large. For example, for $\Omega=0$ the critical
value of $k$
 is estimated to be $k=K_c(0)=4.60366 $ \cite{Geisel}.
 We shall see that the properties of return times change
 abruptly at $k=K_c(P/Q)$, and can be used to
 estimate $K_c(P/Q)$.


\section{Return times for sub-critical and  critical dynamics}

The numerically observed return times  for $k<1$ and diophantine
rotation numbers $\omega$ are in agreement with
 a straightforward extension of Slater's theorem.
The same result is found at the critical value $k=1$.
 For generic interval and each quadratic irrational
 $\omega$ that we have studied, there are again
 only three return times, one is the sum of the other two and two
  always coincide with the
 denominators of the corresponding two successive approximants of $\omega$,
 see for example figure 1a.
 We are fairly confident that
 the conclusions are valid for any quadratic $\omega$ and the maps in the
 class (5),
 but we can not claim anything for homeomorphisms
 with a generic rotation number (see the reference \cite{Chuelo}).

 The three return times,   and their relative weights depend on the
 location and the size of the interval.
 However, there is a sequence of intervals, obtained by partitioning the circle
 with the  iterates  of an  initial point $\theta_0$
that is best suited for the analysis of the return times at a
point $\theta_0$ of
 the map  with a given irrational rotation number $\omega$.
 One considers   the trajectory   formed by  the $q_{i-1}+q_{i}-1$
 successive iterates of  a point  $\theta_0$,
 where $p_i/q_i$ are rational continued fraction
 approximates of the   $\omega$.

  The $q_i$-th and the $q_{i-1}$-th iterate
 form the boundary of an interval $A_i(\theta_0)$ which contains only the initial
 point $\theta_0$ and no other points of the considered part of the trajectory.
 The points generated
 by $q_{i+1}+q_{i}-1$ iterates will subdivide the intervals generated
by  $q_i+q_{i-1}-1$ iterates. One obtains a sequence
$A_{i+1}\subseteq A_{i}$ of intervals that converge
 to the point $\theta_0$ on the orbit of $T_{k,\Omega}$.
  Calculating the
 return time $\tau_{A_i}$  for such sequence of intervals
 $A_i,\> i\rightarrow \infty$  is best adopted to the calculation of the
 return time at the point $\theta_0\in \bigcap_{i=1}^{\infty}
 A_i$. Suppose that the  $q_{i-1}$-th iterate is to the left of $\theta_0$
which is to the
 left of the $q_{i}$-th iterate.
Then, each of the points $\theta\in A_i(\theta_0)$ that are on the
 left side of $\theta_0$  have a unique first return time equal to $q_i$, and
 points $\theta\in A_i(\theta_0)$ to the right of $\theta_0$ also have a unique
 first return time equal to $q_{i-1}$. Thus, there are only
 two return times equal to $q_{i-1}$ and $q_i$.
Furthermore, the union of the intervals formed by
$q_{i+1}+q_{i}-1$ iterates
 gives a partition ${\cal P}_i$ of the circle, and the
  return times into various intervals
 of a partition ${\cal P}_i$ are the same $q_i$ and $q_{i-1}$,
 although the relative weights are obviously
  different for a nonlinear map. However, as $\i\rightarrow \infty$, and for an
 irrational number $\omega=[a,a,a,\dots]$ with a constant tail of the continued fraction expansion,
 the relative weights of the two return times become independent of $i$,
 which implies the existence of a point spectrum independent of $\theta$ (see Appendix)
\be
 F(t)=w_{-}\delta (t-t_{-})+w_{+}\delta (t-t_{+}),
 \ee
  where
\be
 t_{-}=\lim_{\i\rightarrow \infty} {q_i\over <\tau_{A_i} >}={1+a+\omega\over 1+2\omega},
\qquad t_{+}=\lim_{\i\rightarrow \infty} {q_{i-1}\over <\tau_{A_i} >}= \omega t_{-},
 \ee
 are the renormalized return times
in the limit and $q_i$ are denominators of the continued fraction
approximants of $\omega$.
  Since the map is smoothly
  conjugated for $k<1$  and diophantine $\omega$ to a linear rotation, the above
  properties follow from a proposition we prove in this case in the  Appendix.

{\sl Approximation of the measure}:
  The distribution of the return times in various intervals
can be illustrated using the  mean return time. In fact, for any
homeomorphism of the circle there is a unique invariant ergodic
measure which is absolutely continuous with
  respect to the Lebesgue measure on the circle, and the density
  of this measure can be obtained using return times,
 as is indicated by the Kac lemma.
  In figures 1b,c,d we plot density of a coarse-grained mean return time, i.e. the ratio
  between uniform average of the return times into an interval of a partition of the circle
   divided by the Lebesque measure of the interval,
   for a sub-critical and the critical sine-circle map.
   For a sufficiently fine partition this quantity illustrates the
   density of the unique invariant ergodic measure for the
   considered map.
The main computational cost of this method is due to the
computational
 time, complementary to the perturbative method where the main requirement is for a
 sufficient storage space. In order to achieve an resolution of $10^{-m}$ by the return times
 calculations one needs roughly $10^{m}\times K 10^{m}\times 10^{2m}$ iterations of the
 map. Here, $10^{m}$ is the number of intervals of the partition, $K 10^{m}$ with $K<10$ is
 roughly the average return time and $10^{2m}$ is the number of points in each of the intervals.
The same space resolution is archived by using $10^{m}$ Fourier
components of the conjugation
 function in the perturbation method. For $1$ percent accuracy, i.e. $m=2$, both methods
 can be easily implemented, but $10^{-3}$ accuracy
 represents a more challenging task for both the
 methods.

  For $k<1$ the density of the invariant measure
 is a smooth function (see figure 1b,c), and for the critical maps the density becomes
 singular
  (figure 1d) \cite{Khanin2},\cite{Khanin3}.
 Fractal properties of the ergodic measures for the critical
circle maps (5) have been studied.  \rm
  There is strong evidence \cite {Khanin3}
that the class of critical  maps with
 the same fractal spectrum of the invariant measure are characterized only by the rotation number (actually
 probably only by the tail in it's continued fraction expansion \cite{Buric1})
 and by the type of the
 singularity that induces the critical behaviour.


\section{Return times in the super-critical cases}

\rm A super-critical map could have chaotic orbits, at most two
stable attracting  orbits and orbits asymptotic to these.
 Our numerical computations
support a conclusion that for any $\Omega$ there is sufficiently
large $k$,
 such that the distribution of the first return times
 is given by exponentially fast decay. Furthermore, there
  are maps $T_{k,\Omega}$ which have chaotic orbits for $k=1+\epsilon$ with arbitrary
  small but non-zero $\epsilon$, and maps $T_{k,\Omega}$ that show
   chaotic behaviour only for sufficiently large $k>K_c(P/Q)$.
   This is manifested by
 two different roots to the exponential decay of the distribution
 of the first return times.

The following three typical situations can be clearly
distinguished by studying the properties of the return times.

\subsection{Quasi-periodic route}

\rm Consider, first, a weakly-super-critical map $(k,\Omega)$ with
$k=1+\epsilon$, where $\epsilon$ is arbitrary small but non-zero,
and with $\Omega$ such that there is an orbit with an irrational
rotation number. For example, suppose that the rotation number
$\omega(\theta_0)$ of the orbit through point $\theta_0=0$ is
numerically equal to the golden mean $\gamma$. We can use the
return times into a sequence of nested intervals that shrinks to
$\theta_0$, or analogous sequences that shrink on other points  to
study the dynamics.

  Results of such analysis are shown in figures 2a,b,c,d
 which illustrate the
    dynamics of the same map $(k_0,\Omega_0)$ but as
    revealed by the statistics of the return times into a sequence
    of nested intervals of decreasing size.
 Here $k_0=1.01$ and
  $\Omega_0=0.606494989$,  leading to an orbit with
  $\omega$ whose continued fraction expansion is
 $\omega=[0,1,1,1,1,1,1,1,1,1,1,1,2,1,\dots]$, and which has denominators of the
 convergents
 $q_i=2,3,5,8,13,21,34,55,89,144,233\dots$.

  The fact that the map
 is not topologically equivalent to a uniform rotation is
 manifested already at the resolution given by
  the interval of finite size. For example, for the interval $(0,0.0015)$ (fig. 2a)
   there are four return times $55,89,144,199$, where the
 fourth $199$ is not a denominator of any of the convergents to $\omega$.
However, it is  the denominator
 of the Farey neighbor $123/199$ of the approximant $89/144$.
 This signals that, for $k=k_0$
 the $89/144$-tongue and the $123/199$-tongue have
common part of their interiors,
 and that the point $(k_0,\Omega_0)$
 belongs to  both tongues.
The interval that will detect the existence of two intersecting
 tongues at $(k_0,\Omega_0)$ must be smaller than the
 distance between $\theta_0$ and its $144$-th iterate. The first return times into a larger
  interval
 are at most 144 for all its points, i.e. with the resolution weaker
 than the points of the first 144 iterates of $\theta_0$
 the map looks as a smooth rotation.

 Further analysis of the return times
 for the same map $(k_0,\Omega_0)$ but
 into smaller intervals reveals intersections of tongues at the
 deeper levels of the Farey tree between $89/144$ and $123/199$.
 This is illustrated in figures 2b,c which show the return
 times $89,144,199,233,288,343$ into the interval $(0,0.001)$,  and a very large number of
   return times into the interval $(0,0.0001)$.
  The statistics of the return times into $A=(0,0.0001)$ is
  illustrated in figure 2d, by plotting the logarithm of the probability density
   $\ln F_{\theta_0}(t)$ of the
  return time larger that $t=\tau/<\tau_A>$ versus the normalized time $t$.
 The distribution is given by exponential decay with an exponent
 that is numerically close to 1. Thus, on the sufficiently small
 scale, the  map has the distribution of the return times characteristic of
  strongly chaotic systems.

 Increasing $k$, and changing $\Omega$ so that there is
  always an orbit with rotation number $\gamma$,  moves the
 point $(k,\Omega)$ into the domain were the intersections of
 tongues, and the chaotic behaviour, can be detected using larger
 intervals.
 Figures 3a,b,c, show the effects of more intersected tongues on the distribution of return times
  into the same interval $A=(0,0.0015)$.  For $k=1.2$,
 the return times and their relative weights are such that the distribution $F_{\theta_0}(t)$
  is given by exponential decay
  with the exponent that is again numerically equal to 1. In fact,
 for any $k>1$ one can find pairs  $(k,\Omega)$ which imply non-periodic orbit
 through the point $\theta_0$, and the the density $F_{\theta_0}(t)$ for such  map $(k,\Omega)$
  is $F_{\theta_0}(t)=\exp(-t)$. This is illustrated in figure 4a for
  $k=1.3;3;4;5$ and $k=6.14$ and the corresponding $\Omega$.
   For all these maps and for all tested $\theta$ the
  distribution $F_{\theta}(t)$ is always
  $F_{\theta_0}(t)=\exp(-t)$.

  Numerical evidence, presented in figures 2 and 3, suggests the
    following conjecture: Suppose that a point $\theta_0$ lies on
    a quasi-periodic or on a chaotic orbit of a map
    $(k=1+\epsilon,\Omega)$, for arbitrary $\epsilon\neq 0$ and consider a sequence $A_i$
      of nested      intervals  containing  $\theta$
   whose length approaches zero  by increasing $i$.
 The number of different return times  also
    increases with $i$,  and asymptotically,  as $A_i$ shrink  to
    $\theta$,  the distribution of return times becomes
    continuous. Furthermore, the density of probability
  $F_{\theta}(t)$  with respect
    to the Lebesgue measure on the circle of the normalized first return time
     larger than $t$ is given by exponential decay.

\subsection{(Bi)-stability}

 \rm  Possible existence of attracting periodic orbits can be
detected
  by studying the return times into various intervals of a single
   partition of the circle. Although there could be no bounded
   invariant density in this case the return times are still well
   defined.
  The return times into different intervals depend on whether the stable
    orbit have points in the considered interval or not.
   If there is an  attracting periodic orbit with no points in the
   interval than, obviously, there are points in the interval that
   will never come back, i.e. with the first return time $\tau=\infty$,
    indicating the existence of the attracting
    orbit. In this case there will be only those
   return times that correspond to orbits that re-visit the
   interval at least once before being attracted to the attracting
   periodic orbit.  On the other hand, for
 examples of chaotic maps, illustrated in the figure 4a, the return time
 statistics for increasingly fine partitions
  confirms that these maps have no attracting periodic orbits.

\subsection{Period-doubling route}

\rm

In order to study how the period-doubling route to an ultimately
chaotic map $(k,\Omega=P/Q)$ is manifested in the properties of
the return times,  consider the maps $(k,\Omega=0)$ for various
$k$. Results are illustrated by various curves in figure 4.
 The return times into intervals
 at $\theta_0=0$ for any $k\in(1,K_c(\Omega=0)$ show the existence
 of stable periodic orbits, as described in the previous
 subsection. Suddenly, at the accumulation points of the
  period-doubling cascade $k=K_c(\Omega=0)$, the distribution of
  the return times becomes continuous. For such critical $k$, the
  distribution $F_{\theta_0}(t)$ at the point $\theta_0=0$ is given by the
   exponential decay $a\exp(-\alpha t)$, with
   $a<1$ and $\alpha\approx 0.8\neq 1$. Furthermore, still for $k=K_c(\Omega=0)$
   the return times into intervals at $\theta_0$ of finite size are given by
   double exponential curves, which converge to the single
   exponential $\exp(-\alpha t)$ as the intervals shrink to
   $\theta_0$.
   For example, for the interval $A=(0,0.01)$ the
   distribution $F_A(t)$ is well approximated by
    \be
 F_A(t)=0.15\exp(-0.257t)+0.7\exp(-1.8t)
 \ee
 represented by the dash-ed line in figure 4b.

 Other curves in figure 4 represent the distributions of the return
 times for examples of strongly super-critical maps, and
  for intervals of decreasing size around different points.
  The curves with the unique slope in figure 4b are the distributions for maps in  the tongues
   with $\Omega=0$ for $k$ beyond the accumulation of
    period doublings, i.e. for $k>4.604$
   and with $\Omega=1/2$ for $k>1.978$. In all the cases, the
   distribution is given by exponential decay that converges to $F_{\theta_0}(t)=a\exp(-\alpha t)$, where
   $\alpha\neq 1$  and $\theta_0$ is  on an unstable
   periodic orbit. Non-linear curves in figure 4b represent the
 distributions for the map at the accumulation of period doublings and
 on the indicated intervals of decreasing size.
 The curves in 4a represent  $F_{\theta}(t)$
 for examples of strongly super-critical maps at $\theta$ not on a periodic
    point, when $F_{\theta_0}(t)=\exp(-\alpha t)$ with $\alpha\approx 1.0$. Shown are examples with
  $k=1.3,3,4,5, 6.14$ and the corresponding $\Omega$ as explained
   earlier, and also examples  of maps with $\Omega=0;1/2$
   and large $k$.
    All this is consistent with the
   statistics of the return times for other examples of strongly
   chaotic maps.

\section{ Summary and conclusions }

We have analyzed the circle map numerically  for a wide range of
values in the parameter space $(k,\Omega)$ and different sets of
initial conditions. The results can be summarized as follows.

{\sl Sub-critical and critical region}:  For $k< 1$   three
return times are observed and this is theoretically and numerically justified. For
diophantine rotation numbers $\omega$  this result can be
presented as a corollary of Slater's theorem,  since the map is
diffeomorphic to a linear rotation, to which such a theorem
applies. For the case of a special sequence of nested intervals
including a given  point we provide a very simple proof of
Slater's theorem, showing that there are  only two return times
and proving the existence of a limit point spectrum  $F(t)$. The
critical dynamics $k=1$ is also clear since  three return times
for a generic interval and for each quadratic irrational rotation
number are observed. For $k\leq 1$ and $\omega$ diophantine  a
piece-wise constant approximation to the invariant measure is
obtained from the average return times from a uniform partition
of the circle with a very simple procedure and  an accuracy
comparable to other methods.

 {\sl Super-critical region}:
 The dynamics of a map $(k,\Omega)$ in the weak super-critical
 case $k=1+\epsilon$ is dictated
 by the $P/Q$ tongues with a nonempty intersection that contains
 the point $(k,\Omega)$.  This is manifested, and could be detected, in
 the distribution of the first return times
 by appearance of more than three return times, which correspond to the
 rationals on the Farey tree in-between the
  the two major overlapping tongues that contain $(k,\Omega)$. In
  the case that the interval contains a point on a non-periodic
  orbit than there is a sub-interval such that the distribution,
   with respect to the uniform distribution
   of initial points, of
  the return times into this sub-interval is typical for strongly
  chaotic systems, i.e. the exponential decay with exponent equal to unity.

\noindent In the intermediate region  two return times or the
exponential-one spectrum  are observed depending on the existence
of attracting periodic orbits. The  way   $F_A(t)\to e^{-t}$
when  the size of $A$ containing $\theta_0$ approaches $0$ depends
on  $(k,\Omega,\theta_0)$  and it is convenient to distinguish
various  routes.

 {\sl Quasi -periodic route}:  For any
non-periodic point $\theta_0$
 of the map $(k,\Omega)$
 with $k\geq 1+\epsilon$  where $\epsilon>0$ is arbitrary small,
  the spectrum is exponential-one.  Choosing a finite
interval $A$ the  spectrum $F_A(t)$ appears as continuous for a
 sufficiently small  $A$.

{\sl Bi-stability}: For $k=1+\epsilon$  there are   values of
$(k,\Omega)$ such that the map has  attracting periodic orbits.
For any  interval $A$ not intersecting one of these orbits one of
the
 return times is $\tau=\infty$, since
many points do not return being attracted by the periodic orbit.
For $1<k<2$  the attractive periodic orbits are present for most
values of $\Omega$.

{\sl  Period-doubling route}: For any rational $\Omega=p/q$  there
is a critical $k_c=k_c(p/q)$ corresponding to the   accumulation
of period-doublings in every tongue. The transition from
(bi)-stability to chaoticity is manifested  abruptly in the
spectrum
 $F_{A}(t)$. For $k=k_c$ the
spectrum $F_A(t)$ is continuous  and  there are intervals such
that it can be fitted with a double exponential: in the limit
$\mu(A)\to 0$ the spectrum  becomes exponential.

 {\sl Strongly
super-critical region} For $k\gg 1$   the map is chaotic and the
periodic orbits are unstable. The spectrum $F(t)$ is
exponential-one except for a set of points of measure zero
corresponding to unstable periodic orbits where the exponential
decay speed is different from 1.  The results for the
super-critical dynamics  indicate that the analysis of  the
return times spectra  in the super-critical case could be a
useful tool  for a  better understanding of the transition from
the weakly to the strongly chaotic regime. It would be
interesting to follow in details the pattern of intersections of
tongues and period doubling bifurcations inside each tongue,
leading to the strongly super-critical case,  by using the return
times spectra and the way they are  approached when a sequence of
nested intervals squeezing to a point is considered.  To conclude
the computation of the return times spectrum is a simple  and
effective way to explore a dynamical system  and  its
bifurcations.

\vskip 1cm
\bf Acknowledgements \rm

N.B. would like to acknowledge worm hospitality of the Department
of Physics of the University of Bologna, and INFN for financial
support.

\section{ Appendix}

 Letting $R_{\omega}$ be the linear map conjugated to $T_{k,\Omega}$
 and $\Theta_0=\Phi^{-1}(\theta_0)$  be the image of the initial point $\theta_0$ we
 consider a partition of the circle $R^n_\omega\Theta_0$  to which
 corresponds  another partition $ T^n_{k,\Omega}(\theta_0)$.
 The order in these partitions is preserved since the maps are diffeomorphic for
 $k<1$ and $\omega$ diophantine
Let the continued fraction expansion of
 $\omega$ be given by
 \be
 \omega=[a_1,\,a_2,\ldots\,]=
 {1\over \displaystyle
a_1+{\strut 1 \over \displaystyle a_2 + \ddots}}
 \ee
and  $p_i/q_i=[\,a_1,\ldots,a_i\,]$ be
 the corresponding rational approximations of order $i$.
The odd and even approximants are  upper and lower
 bounds to $\omega$, converging monotonically to it.
The following recurrence relations hold \be
p_{i}=a_i\,p_{i-1}\,+\,p_{i-2} \qquad \qquad
q_{i}=a_i\,q_{i-1}\,+\,q_{i-2} \ee
and the  following inequalities hold
  \be
 {1\over q_{i-1}+q_{i} }\leq p_{i-1}\,- \omega\,q_{i-1} \leq {1\over q_{i}}
 \qquad
  {1\over q_{i}+q_{i+1} }\leq q_{i}\,\omega-p_{i}
  \leq {1\over q_{i+1}}
  \ee
  Denoting the linear map iterates of a point $\Theta_0$ by
  \be
  \Theta_0^{q_i}\,=\,R^{q_i}_\omega\,\Theta_0=\Theta_0+q_i\,\omega\,\hbox{mod}\,\, 1=
  \Theta_0\,+\,q_i\,\omega-p_i
  \ee
  the   odd and  even sequences   $\Theta_0^{q_{2i-1}}$ and
  $\Theta_0^{q_{2i}}$ converge monotonically from below and
  from above to  $\Theta_0$.  The intervals
  $A_{i}(\Theta_0)=[\,\Theta_0^{q_{i-1}},\,\Theta_0^{q_{i}}\,]$ if $i$ is even
 and $A_{i}(\Theta_0)=[\,\Theta_0^{q_{i}},\,\Theta_0^{q_{i-1}}\,]$ if $i$ is odd,
  form  a  nested sequence of intervals
   $ A_1\supset\ldots A_i\supset A_{i+1}\ldots$
   squeezing to   $\Theta_0$ as $i\to\infty$  and from the previous
 inequalities the following inclusions hold.
  \be
 \left [ \Theta_0-{1\over q_{i-1}+q_{i}},\,
  \Theta_0+{1\over q_{i}+q_{i+1}} \right ] \subset A_i(\theta_0)\subset  \left [
  \Theta_0-{1\over q_{i} },\, \Theta_0+{1\over q_{i+1}}  \right ]
  \ee

  The intervals $B_i=\Phi_{k,\Omega}^{-1}(A_i)=[T_{k,\Omega}^{i-1}(\theta_0),
  T_{k,\Omega}^{i}(\theta_0)]$  enjoy the same properties for any
  diophantine $\omega$ since the map is orientation preserving and
  $\Phi_{k,\Omega}$ is a diffeomorphism. The sequence $B_i$ is a
  nested monotonic sequence of intervals approaching $\theta_0$
  exponentially fast.  According to Kac's lemma the average return
  times for the intervals $B_i$ and $A_i$ are given by the inverse of
  their length, which increases to $0$ exponentially fast with $i$.

  According to  Slater's theorem  for a generic interval
  and a linear map  with  irrational $\omega$, there are three return
  times, the last one  being the sum of the first two.
  The sequences of nested intervals $[\Theta_0^{q_{i-1}},\, \Theta_0 ]$
  and $[\Theta_0, \Theta_0^{q_{i}}]$ enjoy this property.
  For the intervals $A_i$ defined above  the return times are only two and
  we give a sketch of the proof since it
  quite simple.

 {\bf Proposition}: The return times  in the interval
   $A_i(\Theta_0)$  for the linear map with
   an irrational $\omega$ are  two. If $i$ is even (odd) 
   and for $\Theta\in A_i(\Theta_0)$ the return time is $q_{i}$ ($q_{i-1}$) if
   $\Theta<\Theta_0$ and $q_{i-1}$ ($q_i$) if   $\Theta>\Theta_0$.

\medskip\noindent
 {\bf Proof} The proofs for even or odd $i$ are analogous so we consider only the
 even $i$. In order to prove the results we suppose
$\Theta_0$ is sufficiently
 far from the identified ends 0 and 1 that the order preserving relation
becomes the inequality between real numbers. For initial
conditions near 0, the torus defined as the interval $[-1/2,1/2]$
with identified ends should be considered to have the same
correspondence.  The  Lagrange's theorem  states that in the
interval   $1\leq n\leq q_{i}+q_{i-1}-1$  the minimum of
$|n\omega-m|$, where $m=[n\omega]$ is the integer part, is reached
 for $n=q_i$.
 As a consequence, the two points  $\Theta_0^n$ closest to $\Theta_0$
 for $n\leq q_i+ q_{i-1}-1$    correspond to $n=q_{i-1},\,q_{i}$.
 We consider  the sequence of  nested intervals
 $ A_i(\Theta_0)=[\Theta_0^{q_{i-1}}, \Theta_0{q_{i}}]$ and for a fixed $i$ a point
 $\Theta\in A_i(\Theta_0)$  we denote by $\Theta^n=R_\omega^n \Theta$
 the points of its orbits  and examine two possible cases.

 {\bf Case 1 $\qquad$} If $\Theta<\Theta_0$ then:   $\Theta ^n\not\in A_i$ for $n<q_{i}$,
 $\quad$ $\Theta^{q_{i}}\in A_i$.

 To prove the first point we notice that

 \be
 \Theta_0^{q_{i}}-\Theta^{q_{i}}=\Theta_0-\Theta>0
 \ee
and \be
\Theta^{q_{i}}-\Theta_0^{q_{i-1}}=\Theta^{q_{i}}-\Theta^{q_{i-1}}-(
\Theta_0^{q_{i-1}} -\Theta^{q_{i-1}})=
\mu_L(A_i)-(\Theta_0-\Theta) >0 \ee since both $\Theta_0,\Theta$
belong to $A_i$. To prove the second property we notice that the
points $\Theta_n$ closest to $\Theta$ for $n<q_{i}$ are met for
$n=q_{i-2},\,q_{i-1}$ and
$\Theta^{q_{i-1}}<\Theta<\Theta^{q_{i-2}}$ and we show that
both are out of $A_i$.

\be \Theta_0^{q_{i-1}}-\Theta^{q_{i-1}} =\Theta_0-\Theta>0 \ee
 To prove  $\Theta^{q_{i-2}}>\Theta_0^{q_{i}}$ we show that
$\Theta^{q_{i-2}}-\Theta\geq\mu_L(A_i)$. Indeed
 $ \mu_L(A_i)= \omega q_{i}-p_{i}-\omega q_{i-1}-p_{i-1}=
\omega (q_{i}-q_{i-1})-(p_{i}-p_{i-1}) $ \be =\omega
q_{i-2}-p_{i-2} +(a_{i}-1)\,(\omega\,q_{i-1}-p_{i-1}) \ee
Consequently we obtain
\be \Theta^{q_{i-2}} -\Theta=\omega q_{i-2}-p_{i-2}=
\mu_L(A_i)+ (a_{i}-1)\,(p_{i-1}-\omega\,q_{i-1})\geq \mu_L(A_i)
\ee the equal sign holding for the golden mean.

{\bf Case 2 $\qquad$}  If
 $\Theta>\Theta_0$ then:   $\Theta ^n\not\in A_i$ for $n<q_{i-1}$,
 $\quad$ $\Theta^{q_{i-1}}\in A_i$.

 The fist point  is proved observing that
 \be
  \Theta^{q_{i-1}}-\Theta_0^{q_{i-1}}=\Theta-\Theta_0>0
 \ee
and \be
 \Theta_0^{q_{i}}- \Theta^{q_{i-1}}= \Theta_0^{q_{i}}-  \Theta^{q_{i}}
+ \Theta^{q_{i}}
-\Theta^{q_{i-1}}=\mu_L(A_i)-(\Theta-\Theta_0)>0 \ee Concerning
the last point we   notice that we already  proved (two equations
above) that  $\Theta^{q_{i-2}}-\Theta>\mu_L(A_i)$  for any
$\Theta\in A_i$. We prove also that
$\Theta-\Theta^{q_{i-3}}>\mu(A_i)$. Indeed from
$\Theta-\Theta^{q_{i-3}}-\mu(A_i)= p_{i-3}-\omega q_{i-3}-
(\omega q_{i}-p_{i}) + (\omega q_{i-1}-p_{i-1})=$

$=p_{i}-a_{i-1}\,p_{i-2}-\omega\,( q_{i}-a_{i-1}q_{i-2})\geq
  p_{i}-p_{i-2}-\omega(\, q_{i}-q_{i-2}) $

$\qquad\qquad\qquad\qquad= a_{i}(p_{i-2}-\omega\,q_{i-1})>0$,
where we have used twice the recurrence relations for the $p_i$
and $q_i$,   taking into account that
$\omega\,q_{i-2}-p_{i-2}>0$, and that $a_i \geq 1$ since we have
assumed $\omega$ to be irrational.

  {\bf Existence of the spectrum}:  The return times in
the intervals $A_i$ are given by $q_i$ if $\Theta\in
A_i^{-}\equiv A\Theta\in [\Theta_0^{q_{i-1}},\Theta_0]$ and
 $q_{i-1}$ if
$\Theta\in A_i^{+}\equiv [\Theta_0, \Theta_0^{q_{i}}]$. As a
consequence in we consider the  return times in $A_i$ the
relative measures of points that come back to $A_i{-}$ and to
$A_i^{+}$ will be given by the Lebesgue measure of these
intervals normalized to the Lebesgue measure of $A_i$. As a
consequence  the   weights $w_-,\,w_+$ for the return times $q_i$
and $q_{i-1}$ are given by
\be w_-= {\mu_L(A_i^{-})\over\mu_L(A_i) }= {1\over 1+r} \qquad
w_+= {\mu_L(A_i^{+})\over\mu_L(A_i) }= {1\over 1+r^{-1} } \qquad
r={\mu_L(A_i^{+})\over\mu_L(A_i^{-})} \ee
 We notice that these weights for a quadratic irrational are independent
 of $i$. Indeed in this case $p_i=q_{i-1}$ and letting
 $\omega=[a,a,a,a,\ldots]$
 \be
r={\omega \,q_{i}-q_{i-1}\over q_{i-2}-\omega
\,q_{i-1}}={1\over a+\omega}
 \ee
 Indeed  taking into account that $\omega(a+\omega)=1$,  the above relation
 is verified if  $q_{i}-(a+\omega) q_{i-1}-q{i-2}+\omega q{i-1}=$
 \phantom{.} $q_{i}-aq_{i-1}-q_{i-2}$, which is the case due to the
 recurrence for $q_i$.  As a consequence the average time is
\be \langle \tau_{A_i} \rangle =
 w_-\,q_{i} +w_+\,q_{i-i}={ (a+\omega)\, q_{i} +
  q_{i-1} \over 1+a+\omega}
 \ee
 The normalized times  in the limit $i\to \infty$ become
$$  t_-= \,\lim_{i\to \infty}{ q_i\over \langle \tau_{A_i} \rangle}=
 (w_++\omega w_+)^{-1}={1+a+\omega\over a+ 2\omega}$$
 \phantom{.}
 \be
 t_+= \,\lim_{i\to \infty}{ q_{i-1}\over \langle \tau_{A_i} \rangle}=
 (w_-+\omega w_+)^{-1}={\omega\,( 1+a+\omega)\over a+ 2\omega }=\,\omega\,t_-
 \ee
 As a consequence the limit point spectrum exists and is given by
 \be
 F(t)=w_- \,\delta(t-t_-)+w_+ \,\delta(t-t_+)
 \ee
 We believe that the spectrum exists for any irrational $\omega$.
 The existence of a spectrum for a generic nested sequence
  of intervals, squeezing to $\theta_0$, remains an open question.

\newpage

{\bf FIGURE CAPTIONS}

\bf Figure 1a,b,c,d:\rm  Illustrate the sub-critical and critical
cases with $\omega=\gamma$ :(a)Three return times for
$k=1;\Omega=0.606661; A=(0,0.0015)$, and densities of the unique
invariant ergodic measure for (b) $k=0.75, \Omega=0.61088669$ (c)
$k=0.9, \Omega=0.6083938$ and (d) $k=1, \Omega=0.606661$.

\bf Figure 2a,b,c,d: \rm Illustrate the distribution of the return
times into a sequence of sub-intervals for the fixed weakly
super-critical map $k=1.01; \Omega= 0.606494989$.

\bf  Figures 3a,b,c,d:\rm Illustrate the return
 times into the
interval $(0,0.0015)$ for  weakly super-critical maps with
$(k,\Omega)=$ (a) $(1.015,0.6063931)$ (b) $(1.1,0.6054765)$ (c)
$(1.2,0.6047099)$ (d) shows $\ln F_{\theta}(t)$.vs.$t$ for
$k=1.2,\Omega=0.6047099$, with the slope $0.9996$.

\bf Figures 4a,b: \rm Strongly super-critical dynamics: (a) $\ln
F_{\theta}(t)$.vs.$t$ with slope $\approx 1.0$ for
$(k,\Omega)=(1.3,0.606187),(3,0.51738887)$,$(4,0.36176849)$,$(5,0.20197433)$,
$(6.14,0.7184975)$ and for
$(k,\Omega)=(4.605,0),(2\Pi,0)$,$(1.98,1/2),(2,1/2)$;

 b)$\ln F_A(t)$.vs.$t$ with slope $\neq 1$,  for $(k,\Omega)=(4.60366,0)$ when $A=(0.01)$;
 $(0,0.0005)$ and
 $A=(0,0.0001)$, with approximation (14) (dash-ed line).
 Other curves represent $\ln F_A{\theta}(t)$.vs.$t$  at unstable periodic
points for other examples with $\Omega=0;1/2$. The maximal slope
is $0.85$.


\vskip 1cm

\begin{thebibliography}{99}

\bibitem{Chir99}{B.V. Chirikov and D.I. Shepelyansky, Asymptotic Statistics of Poincar\' e
 recurrences in Hamiltonian Systems with Divided Phase Space, {\it Phys.Rev.Lett.} {\bf
82}: 528, (1999). }

\bibitem{Vaienti1}{M. Hirata, B. Saussol, and S. Vaienti, Statistics of return times:
 a general framework and new applications, Commun.Math.Phys. {\bf 206}
: 33, (1999).}

\bibitem{Ors}{S. Ornstein and B. Weiss,
 Entropy and data compression, {IEEE Trans.inf.Theory} {\bf 39}: 78 (1993).}


\bibitem{Vaienti2} {N. Haydn, J. Luevano, G. Mantica and S. Vaienti, Multifractal
 properties of return time statistic,  preprint, (2001).}

\bibitem{Hir95}{M. Hirata, Poisson law for dynamical systems with the "self-mixing"
 condition, in {\it Dynamical systems and chaos}
 {\bf 1}: 87 (World Sci. Pub., New York, 1995).}

\bibitem{transp}{J.D. Meiss, Simpectic Maps, variational principles and transport,
 {\it Rev. Mod. Phys.} {\bf 64}: 795 (1992).}


\bibitem{Young}{L.S. Young, Recurrence times and rates of mixing,
 {\it Israel J. of Mathematics}, {\bf 110}: 153, (1999).}

\bibitem{Aldo}{A. Rampioni, G. Turchetti and  S. Vaienti,
Contribution to  the summer
 school "Mathematical Aspects of Quantum Chaos", Bologna 1-10 Septembre, (2001). }

\bibitem{Hu} {H. Hu, A. Rampioni, G. Turchetti, S. Vaienti {\it
Polynomial law decay of the return times spetrum  for integrable
area preserving maps} in preparation}

\bibitem{Zavlavsky}{G.M. Zaslavsky and M. Edelman,
 Weak mixing and anomalous kinetics along filamented surfaces,
 {\it Chaos} {\bf 11}: 295 (2001). }

\bibitem{us2002}{N. Buric, A. Rampioni, G. Turchetti and S. Vaienti, Poincar\' e Reccurences for Area-Preserving
 Maps,  submitted to
 {\it Phys.Rev.Lett.} (2002). }

\bibitem{Slater}{N.B. Slater, Gaps and steps for the sequence $n\theta$ mod $1$,
  {\it Proc.Camb.Phil.Soc.} {\bf 63}: 1115, (1967).}

\bibitem{Mayer}{H.D. Mayer, On the distribution of reccurence times in nonlinear systems,
 {\it Lett.Math.Phys.} {\bf 16}: 139, (1988).}


 \bibitem{Denj}{ A. Denjoy, Sur le curbes definies par le
 equation diff\' erentieles a la surface du tore, J. Math. Pures. Appl 1932,  {\bf  11}: 333, (1932). }

\bibitem{Herman}{M. Herman,  Sur la conjugasion diff\' erentiable des diff\' eomorphismes du
 cercle a des rotations, Pub.Mat. IHES {\bf 49}: 5, (1979).}

\bibitem{Geisel}{T. Geisel and J. Nierwetberg, Onset of diffusion and universal scaling in chaotic systems,
 {\it Phys.Rev.Lett.} {\bf 48}: 7, (1981).}

\bibitem{Glass}{L. Glass and R. Perez, Fine structure of phase-locking,
 {\it Phys.Rev. Lett.} {\bf 48}: 1772,( 1982).}

\bibitem{Bohr1}{M.H. Jensen, P. Bak and T. Bohr, Transition to
chaos by interaction of resonances in dissipative systems. I.
Circle maps,
 {\it Phys.Rev. A} {\bf 30}: 1960, (1984).}

\bibitem{Bohr2}{T. Bohr and G. Gunaratne, Scaling for supercritical circle
maps: Numerical investigation of the onset of bistability and
period doubling,
  {\it Phys. Lett.} {\bf 113A}: 55, (1985).}

\bibitem{Boyland}{P.L. Boyland, Bifurcations of circle maps: Arnold's
 tongues, bistability and rotation intervals, {\it Commun. Math. Phys.} {\bf 106}: 353, (1986).}

\bibitem{MacKay}{R.S. MacKay and C. Tresser, Transition to topological chaos for circle maps,
 {\it Physica} {\bf D19}: 206, (1986).}

\bibitem{Rand}{D. Rand, S. Ostlung, J. Sethna and E.D. Siggia,
Universal transition from quasiperiodicity to chaos in
dissipative systems, {\it Phys.Rev.Lett} {\bf 49}: 132, (1982).}

\bibitem{Buric2}{B. Buric, J.R.  Cartwright, O. Piro and I.C. Percival,
 On modular smoothing and scaling functions for mode-locking,
  Limit laws for entrance times for homeomorphisms of the circle,
   {\it Phys.Lett.A} {\bf
163}: 63, (1992).}

\bibitem{Tresser}{K.M. Bruks and C. Tresser, A Farey tree
organization of locking regions for simple circle maps,
 {\it Proc. Amer. Math. Soc.} {\bf 124}: 637 (1996).}

\bibitem{Khanin2}{K.M. Khanin, Universal estimates for critical circle maps,
 {\it Chaos} {\bf 1} 181: (1991).}

\bibitem{Khanin3}{B.R. Hunt, K.M. Khanin, Ya. G. Sinai and J. A. Yorke,
Fractal properties of critical invariant curves,{\it J.Stat.Phys.}
{\bf 85}: 261. (1996)}

\bibitem{Buric1}{N. Buric, M. Mudrinic and K. Todorovic,
Universal scaling of critical quasiperiodic orbits in a class of
twist maps, {\it J.Phys.A:Math and Gen.} {\bf 31}: 7847, (1998).}


\bibitem{Chuelo}{Z. Coelho and E.de Faria, Limit laws for entrance
 times for homeomorphisms of the circle, {\it Israel J. of Math.} {\bf 93}: 93 (1996).}


\end{thebibliography}
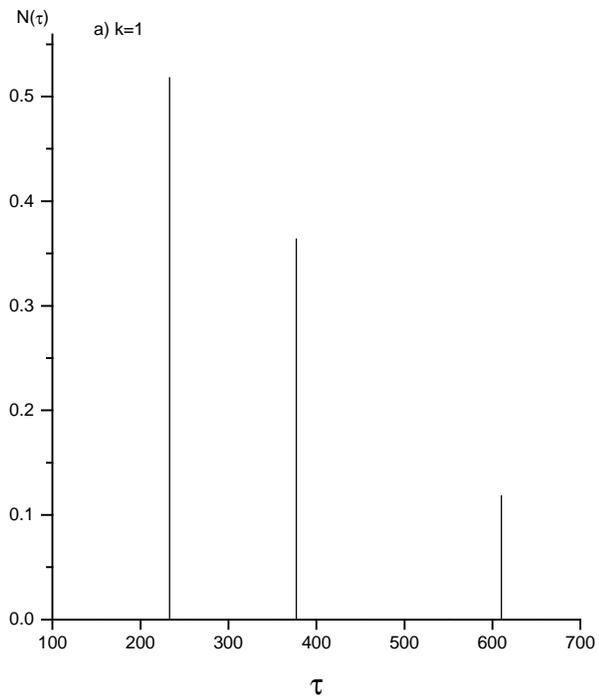
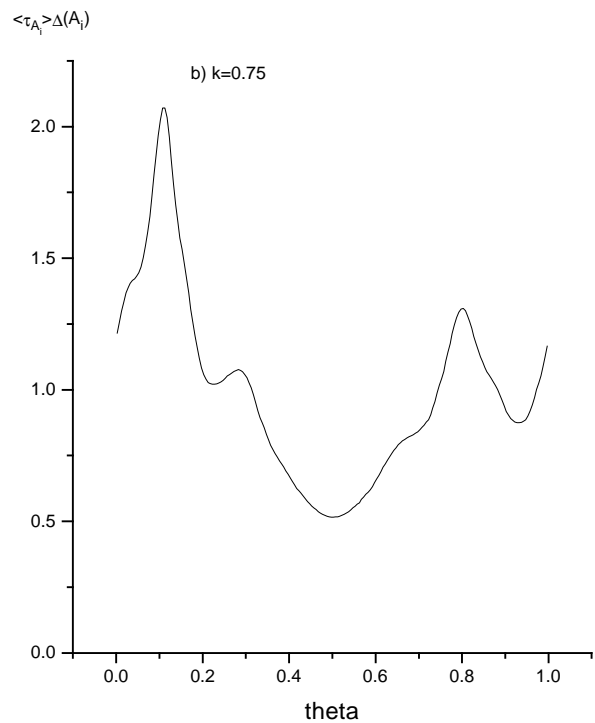
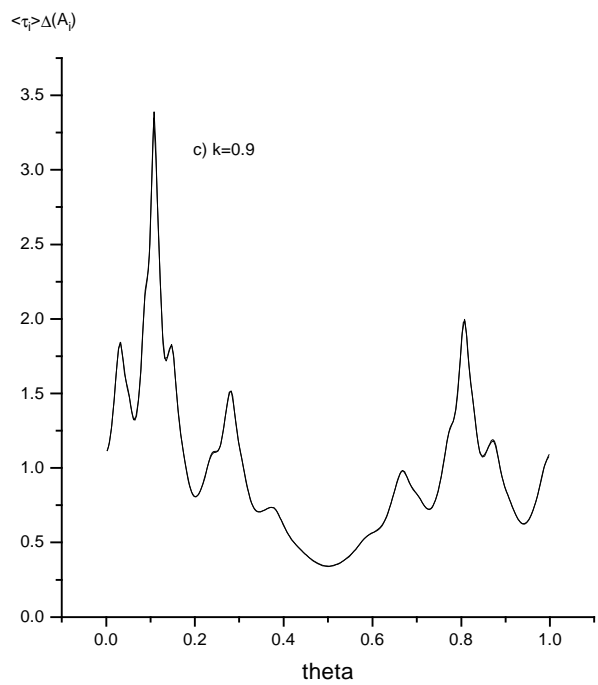
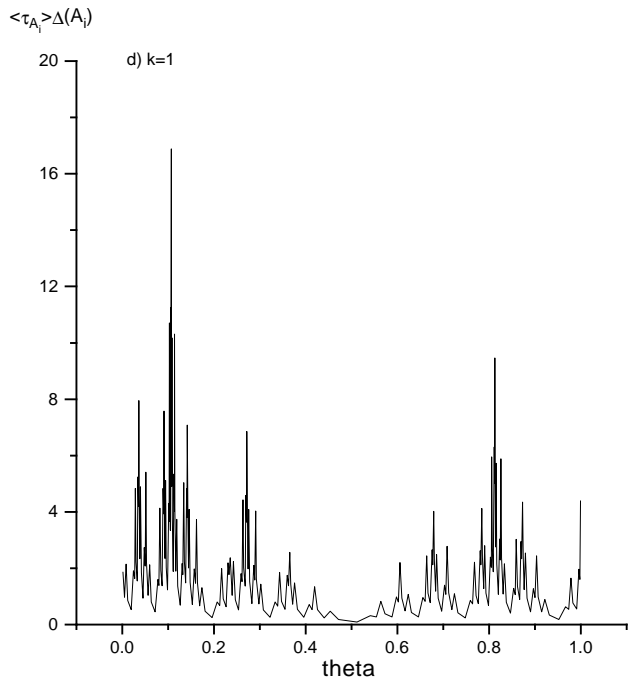
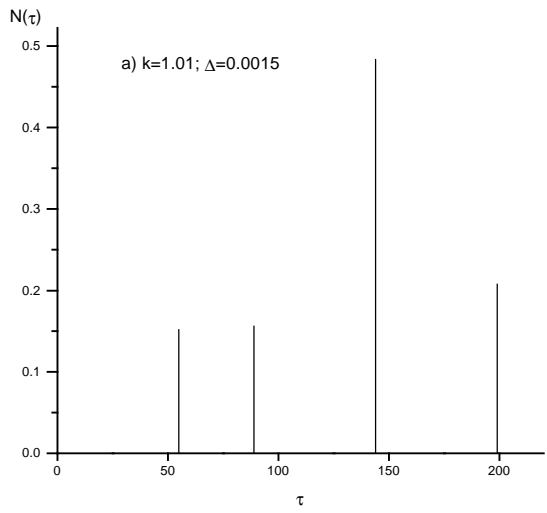
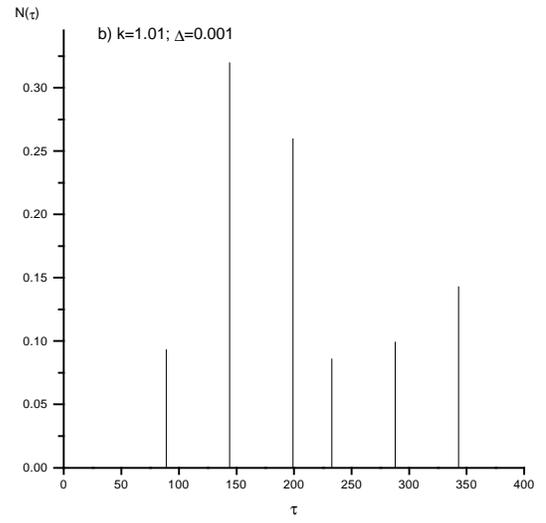
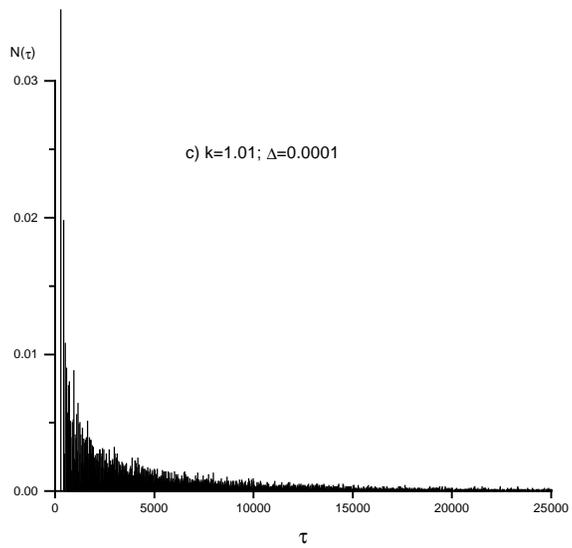
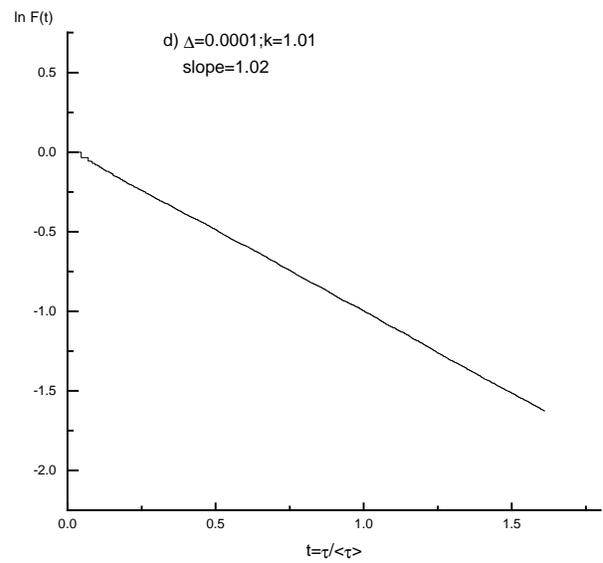
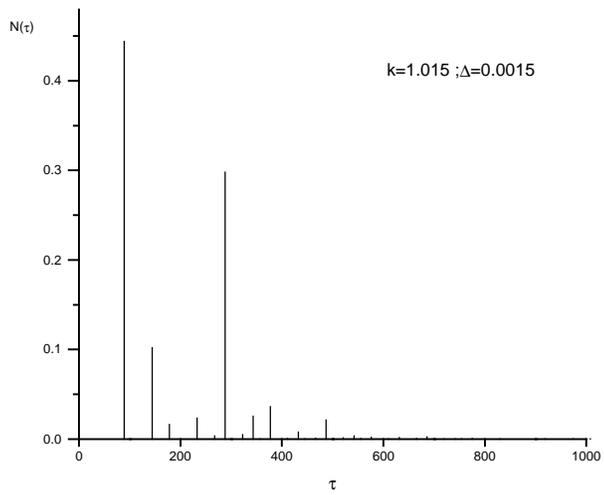
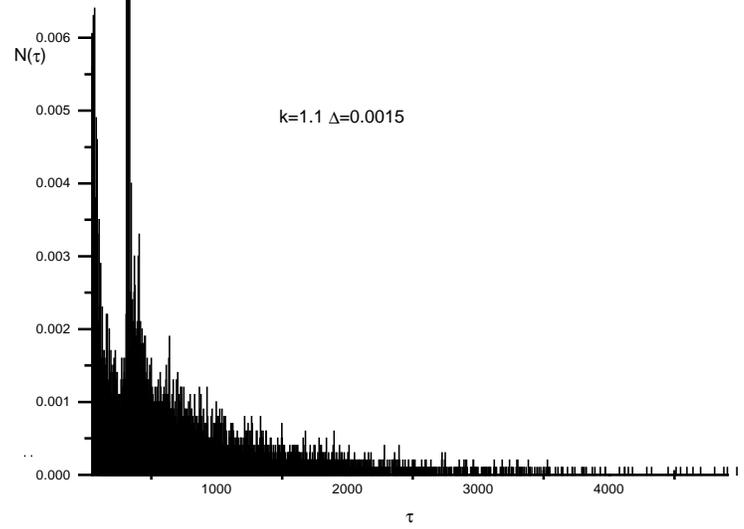
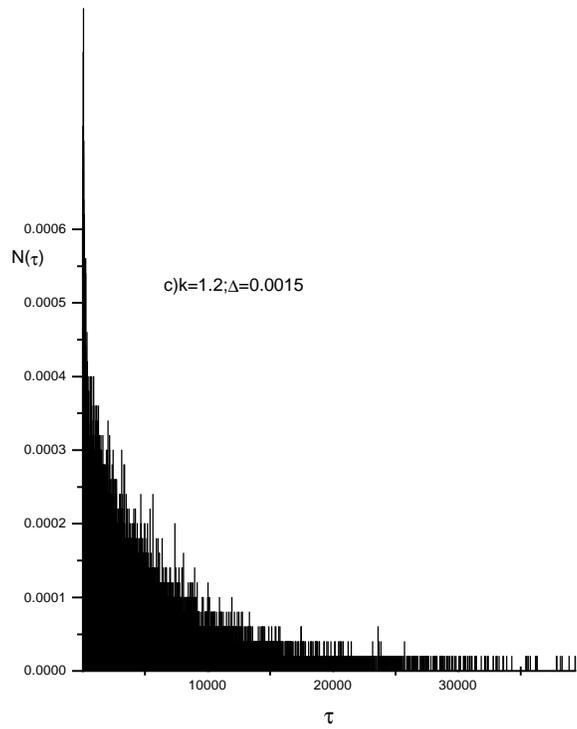
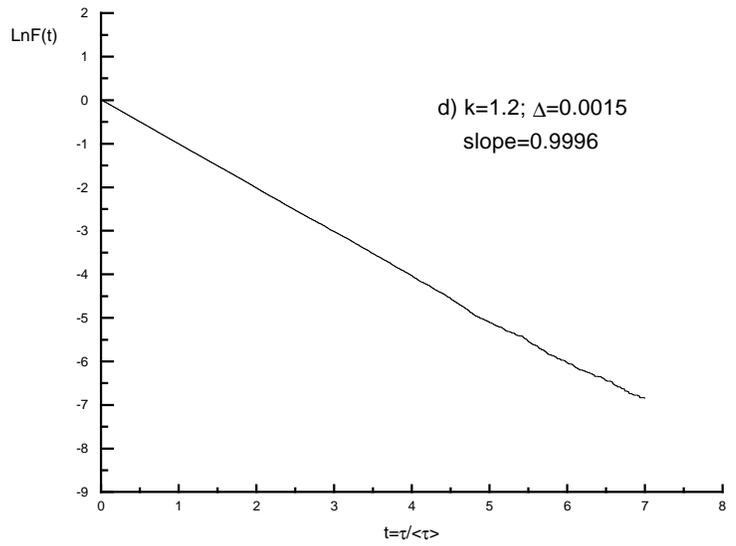
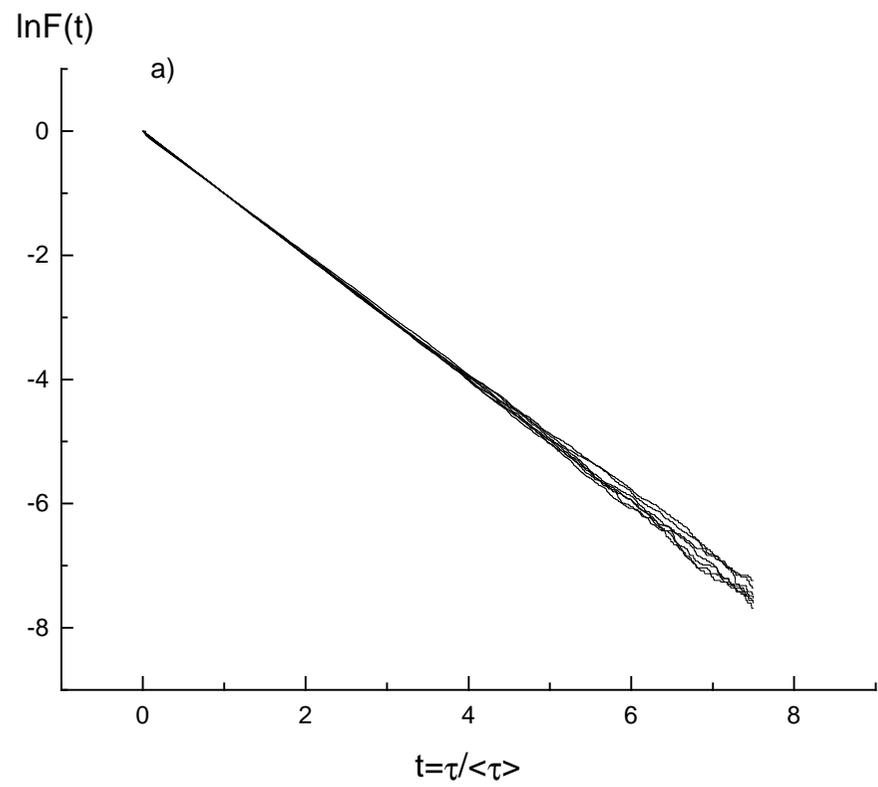
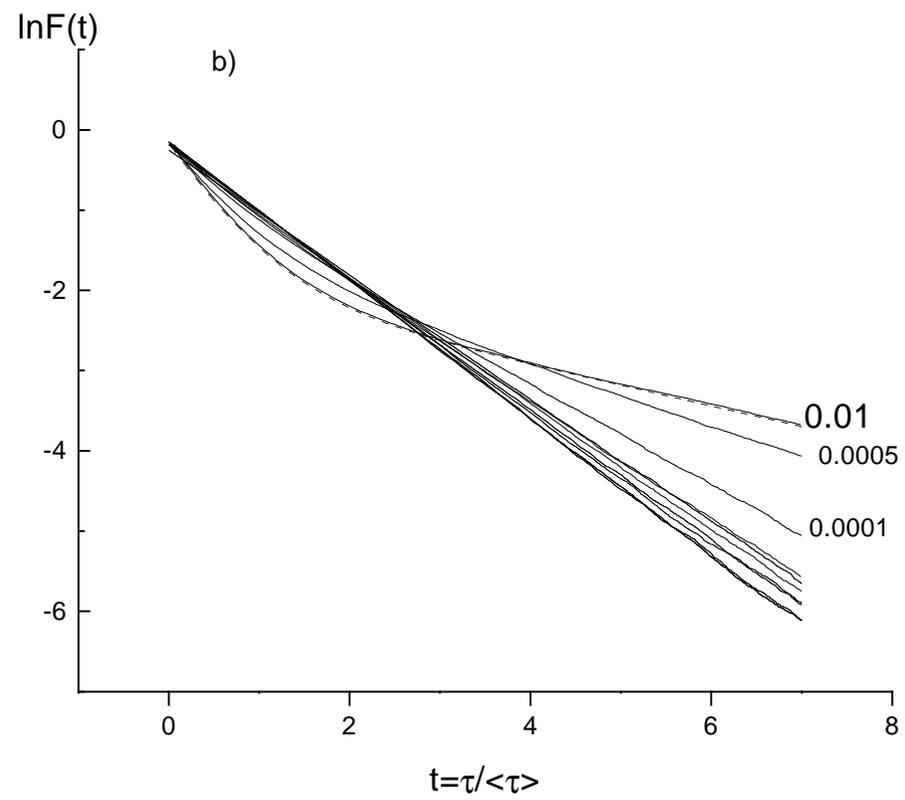
\end{document}